\documentclass[floats,preprint,aps,eqsecnum,nofootinbib]{revtex4}

\newcommand{\smT}{{\rm\scriptscriptstyle T}}
\newcommand{\smS}{{\rm\scriptscriptstyle S}}
\newcommand{\smA}{{\rm\scriptscriptstyle A}}

\global\arraycolsep=2pt                 

\makeatletter
\def\fotnotesize{\@setsize\footnotesize{9.5pt}\xpt\@xpt
\abovedisplayskip 10pt plus2pt minus 5pt
\belowdisplayskip \abovedisplayskip
\abovedisplayshortskip \z@ plus 3pt
\belowdisplayshortskip 6pt plus 2pt minus 2pt
\def\@listi{\topsep 6pt plus 2pt minus 2pt
\parsep 3pt plus 2pt minus 1pt \itemsep \parsep}}
\makeatother

\begin{document}  

\preprint{LA-UR 12-00207}

\title {
    Leading Relativistic Corrections to the Kompaneets Equation
}

\author{Lowell S. Brown and Dean L. Preston}
\affiliation{Los Alamos National Laboratory
\\
Los Alamos, New Mexico 87545}

\begin{abstract}

We calculate the first relativistic corrections to the Kompaneets
equation for the evolution of the photon frequency distribution
brought about by Compton scattering.  The Lorentz invariant Boltzmann
equation for electron-photon scattering is first specialized to
isotropic electron and photon distributions, the squared scattering
amplitude and the energy-momentum conserving delta function are each
expanded to order $v^4/c^4$, averages over the directions of the
electron and photon momenta are then carried out, and finally an
integration over the photon energy yields our Fokker-Planck
equation. The Kompaneets equation, which involves only first- and
second-order derivatives with respect to the photon energy, results
from the order $v^2/c^2$ terms, while the first relativistic
corrections of order $v^4/c^4$ introduce third- and fourth-order
derivatives. We emphasize that our result holds when neither the
electrons nor the photons are in thermal equilibrium; two effective
temperatures characterize a general, non-thermal electron
distribution.  When the electrons are in thermal equilibrium our
relativistic Fokker-Planck equation is in complete agreement with the
most recent published results, but we both disagree with older work.

\end{abstract}

\maketitle

\newpage

\section{Introduction}

The Kompaneets \cite{kom} equation,
\begin{equation}
\frac{\partial}{\partial t} \, f(t,\omega) = 
      \frac{\sigma_\smT \ n_e}{m_e \, c} \, \frac{1}{\omega^2} \,
\frac{\partial}{\partial \omega} \, \omega^4 \, 
\Bigg\{ T \, \frac{\partial \, f(t,\omega)}{\partial \omega} 
    + \hbar \Big[ 1 + f(t,\omega) \Big] \, f(t,\omega) \Bigg\} \,,
\label{komp}
\end{equation}
describes the scattering of unpolarized, low energy photons of 
frequency $\omega$ on a dilute distribution of non-relativistic electrons 
when all the particles --- photons and electrons --- are distributed 
isotropically in their momenta. The non-relativistic total 
photon-electron cross section is the Thomson cross section $\sigma_\smT$.
The electron number density and mass are denoted by $n_e$ and $m_e$.
The photon phase space distribution $f(t,\omega)$ is normalized such 
that the number $n_\gamma$ of photons per unit volume is given by
\begin{equation}
n_\gamma(t) = 2 \int \frac{(d^3{\bf k})}{(2\pi)^3} \, f(t,\omega) \,,
\label{pno}
\end{equation}
in which the prefactor 2 counts the number of photon polarization 
states and ${\bf k}$ is the photon wave-number vector with
$|{\bf k}| c = \omega$. If the electrons are in thermal equilibrium 
described by a Maxwell-Boltzmann distribution, then
$T$ is the temperature (in energy units) $T_e$ of this thermal distribution. 
However, the Kompaneets equation (\ref{komp}) holds for any isotropic 
distribution of electron momenta with $T$ defined to be 2/3 of the 
average energy in this distribution \cite{brown}.  For photons 
with a Planck distribution,
\begin{equation}
f(t,\omega) \to f^{(0)}(\omega) = 
\frac{1}{\exp\{ \hbar \omega \, / \, T_\gamma \} -1} \,.
\label{Planck}
\end{equation}
The terms in the curly braces in the Kompaneets equation (\ref{komp}) 
vanish when $T_\gamma = T$. In particular, if $T = T_\gamma = T_e$,  
there is a time-independent photon distribution in thermal 
equilibrium with the electrons\footnote{Since Compton scattering preserves the
photon number, the collision term on the right-hand side of 
Eq.~(\ref{komp}) also vanishes for a general Bose-Einstein distribution
of massless particles at temperature $T$, 
$ f(t,\omega) \to 
f^{(\alpha)}(\omega) = [ \exp\{(\hbar\omega/T) - \alpha\}-1]^{-1}$.}.

Our purpose here is to examine the first relativistic corrections to
the Kompaneets equation. These corrections have been previously
computed by Challinor and Lasenby (C\&L) \cite{C&L} for the case in
which the electrons are in a thermal distribution. Using the method 
of C\&L, Itoh, Kohyama, and Nozawa \cite{itoh} carried out the expansion 
to a much higher order  in $v/c$. Subsequently, Sazonov and Sunyaev
\cite{SS} confirmed the previous work of Challinor and Lasenby. Here
we use a method that is quite different from that employed by C\&L, a
method that does not require that the electrons be in thermal
equilibrium.  Although the structure of our result is quite different
from that found by C\&L, we agree with C\&L in the number of
higher-order derivatives with respect to the photon frequency $\omega$
which must supplement the Kompaneets equation to correctly account for
the relativistic corrections. Such higher-order derivative terms are
missing from the ad-hoc treatments of Cooper \cite{coop} and of
Prasad, Shestakov, Kershaw, and Zimmerman \cite{prasad}.  These
authors assume (incorrectly) that the relativistic corrections may be
accounted for by simply replacing the factor $\omega^4$ that stands
just before the curly braces in Eq.~(\ref{komp}) by a function
$\alpha(\omega, T)$ which is determined so as to give the rate of
change of the photon energy density including the first relativistic
corrections. We compute both the rate of energy exchange between the
photons and electrons and the Sunyaev--Zel'dovich effect
\cite{ZS,SZ1,SZ2} which follow from the relativistically corrected
Kompaneets equation. Including the first relativistic corrections, our
results entail two effective temperatures $T_{\rm eff \, 1}$ and
$T_{\rm eff\, 2}$ which are defined by energy moments of the electron
phase-space distribution.  When the electron distribution is
restricted to a thermal, relativistic Maxwell-Boltzmann distribution
at temperature $T$, $T_{\rm eff \, 1} = T_{\rm eff \, 2} = T$ and we
find, after some algebra, that our results that have a completely
different structure are, in fact, in complete agreement with those of
C\&L.  Moreover, the rate of energy exchange that we compute (also
written down by C\&L) agrees with that found earlier by Woodward
\cite{wood}.

Our presentation is organized as follows: After describing the 
general method we use in Sec.~\ref{general}, we then outline 
the calculation in Sec.~\ref{outline} using the results of 
several Appendices.  Finally, our results are shown in 
Sec.~\ref{results}: Sec.~\ref{A} presents our general result,
Sec.~\ref{B} gives its restriction to the case in which the photons
are in thermal equilibrium at temperature $T_\gamma$, Sec.~\ref{C} 
derives the rate of energy transport between photons at temperature
$T_\gamma$ and the electrons in a general distribution, and finally, 
in Sec.~\ref{D} the Sunyaev-Zel'dovich effect for non-thermal electrons
with the first relativistic correction is briefly described.

\section{Relativistic Boltzmann Equation for Isotropic Scattering}
\label{general}

We start from the Lorentz invariant form of the Boltzmann equation for 
electron-photon scattering:
\begin{eqnarray}
k\partial f(x,k) &=& \int 
   \frac{(d^3 {\bf p}')}{(2\pi)^3} \frac{1}{2E'} \
   \frac{(d^3 {\bf k}')}{(2\pi)^3} \frac{1}{2\omega'} \,
        \frac{(d^3 {\bf p})}{(2\pi)^3} \frac{1}{2E} \,
    (2\pi)^4 \, \delta^4( p' + k' - p - k )
\nonumber\\
&&
    \Big|T(p',k'; p,k) \Big|^2 \, 
  \Big\{ [ 1 + f(x,k) ] \, 2 \, g(x, p') \, f(x,k') -
          [ 1 + f(x,k') ] \, 2 \, g(x,p) \, f(x,k) \Big\} \,.
\nonumber\\
&&
\label{boltzmann}
\end{eqnarray}
Here we revert to units in which $\hbar = 1 = c$, but we shall return to
conventional units when we write the final result. The left-hand side
of the equation involves the relativistic scalar 
$k\partial = \omega (\partial / \partial t) + {\bf k} \cdot \nabla$. 
We are assuming that the electrons and photons are not polarized. Hence
$|T|^2$ denotes the square of the Lorentz invariant scattering 
amplitude that is summed over the initial and final electron and photon 
spins. It is divided by the initial electron spin weight $g_e = 2$ so as
to describe the average scattering from an initially unpolarized ensemble of 
electrons. It is divided by the square of the photon spin weight 
$g_\gamma^2 = 4$ because initially there is  an unpolarized 
mixture and finally the scattering is into the scalar density $f(x,k)$ 
that describes a typical photon [with the factor $g_\gamma = 2$ needed to
provide the photon number count in Eq.~(\ref{pno})]. The function 
$g(x,p)$ is the electron phase space density. We choose our Lorentz metric 
to have signature $(- \, + \, + \, +)$ so that $t = x^0 = - x_0$ while 
for the spatial coordinates $x^k = x_k$.

We now specialize to the isotropic case of interest where 
$f(x,k) \to f(t,\omega)$ and $g(x,p) \to g(t,E)$, with the electron
number density given by
\begin{equation}
n_e = 2 \int \frac{(d^3{\bf p})}{(2\pi)^3} \, g(t,E) \,.
\end{equation}
The integration variables $p$ and $p'$ in Eq.~(\ref{boltzmann}) are
dummy variables.  We shall make the interchange 
$p \leftrightarrow p' $ in the first set of terms in Eq.~(\ref{boltzmann})
so as to have a common factor of $g(t,E)$ for the two `scattering in to'
and `scattering out of' terms. To keep a convenient form, we shall also
use the detailed balance relation 
\begin{equation}
    |T(p',k'; p,k)|^2  =  |T(p,k; p',k')|^2 
\label{bal}
\end{equation}
for this first term in Eq.~(\ref{boltzmann}).  Finally, we note that
the ${\bf p}'$ integration is best performed using
\begin{equation}
\frac{(d^3 {\bf p}')}{(2\pi)^3} \frac{1}{2E'} =
\frac{(d^3 {\bf p}')}{(2\pi)^3} \frac{1}{2\sqrt{{{\bf p}'}^2 + m_e^2}} =
\frac{(d^4 p')}{(2\pi)^3} \, \delta\left( {p'}^2 + m_e^2 \right)
\end{equation}
against the four-dimensional delta function which now replaces
\begin{equation}
p' = p + k - k' \,,
\label{mom}
\end{equation}
giving
\begin{equation}
{p'}^2 + m_e^2 = 2 p \left( k - k' \right) - 2 k k' \,.
\end{equation}
In this fashion, we obtain
\begin{eqnarray}
\omega \frac{\partial}{\partial t} \, f(t,\omega) &=& \int 
   \frac{(d^3 {\bf k}')}{(2\pi)^3} \frac{1}{2\omega'} \,
        \frac{(d^3 {\bf p})}{(2\pi)^3} \frac{1}{2E} \, 2 \, g(t,E) \,
    2\pi \, \delta\left(  2 p \left( k - k' \right) - 2 k k' \right)
\nonumber\\
&& \qquad
  \Big\{  \Big|T(p',k; p,k') \Big|^2 \, 
  \Big\{ [ 1 + f(t,\omega) ] \, f(t,\omega')
\nonumber\\
&& \qquad\qquad\qquad
 -  \Big|T(p',k'; p,k) \Big|^2 \,
  [ 1 + f(t,\omega') ] \, f(t,\omega) \Big\}  \,,
\label{boltzmann2}
\end{eqnarray}
in which the four-momentum $p'$ in $\Big|T(p',k; p,k') \Big|^2 $ is
determined by Eq.~(\ref{mom}).

The angular part of the integrations over ${\bf p}$ and ${\bf k}'$ pick
out the completely rotationally invariant part of the integrand. Thus, 
with angular brackets denoting the average over all the orientations
of the vectors within it, we may make the replacement
\begin{eqnarray}
 \delta \Big(  2 p \left( k - k' \right) - 2 k k' \Big)
  \,  \Big|T(p',k'; p,k) \Big|^2 &\to&
 \left\langle \delta \Big(  2 p \left( k - k' \right) - 2 k k' \Big)
  \,  \Big|T(p',k'; p,k) \Big|^2 \right\rangle  
\nonumber\\
&\equiv& s(p; \omega',\omega) \,.
\label{ssss}
\end{eqnarray}
In view of these remarks, we may write Eq.~(\ref{boltzmann2}) as
\begin{equation}
\frac{\partial}{\partial t} f(t,\omega) = 
      \frac{1}{\omega^2} \,
        \int \frac{(d^3{\bf p})}{(2\pi)^3} \, \frac{1}{2E} \, 2 g(p) 
              \, F(t, \omega ; p) \,,
\label{colll}
\end{equation}
with
\begin{eqnarray}
F(t, \omega ; p) &=& \frac{\omega}{2\pi} 
\int_0^\infty \!\! \omega' \, d\omega' \, 
\Big\{ s(p; \omega,\omega') \, [ 1 + f(t,\omega) ] \, f(t,\omega') 
  -  s(p;\omega',\omega) \, [ 1 + f(t,\omega') ] \, f(t,\omega) \Big\} .
\nonumber\\
&&
\label{ccolll}
\end{eqnarray}

For the evaluation of Eq.~(\ref{ccolll}) it is convenient to separate
the weight that appears there into symmetric and antisymmetric parts:
\begin{equation}
s(p; \omega',\omega)= s^\smS(p; \omega',\omega) +  s^\smA(p; \omega',\omega) \,,
\end{equation}
with 
\begin{equation}
s^\smS(p; \omega,\omega') = + s^\smS(p; \omega',\omega) \,,
\end{equation}
and
\begin{equation}
s^\smA(p; \omega,\omega') = - s^\smA(p; \omega',\omega) \,.
\end{equation}
With this decomposition, Eq.~(\ref{ccolll}) becomes
\begin{eqnarray}
F(t, \omega ; p) &=& 
\frac{\omega}{2\pi} \int_0^\infty  \omega' \, d\omega' \, 
 \Big[ s^\smS(p; \omega',\omega) - s^\smA(p; \omega',\omega) \Big]
\,  f(t,\omega') 
\nonumber\\
&& \qquad
    -  \frac{\omega}{2\pi} \, f(t,\omega) \, 
\int_0^\infty \omega' \, d\omega' \, 
 \Big[ s^\smS(p; \omega',\omega) + s^\smA(p; \omega',\omega) \Big]
\nonumber\\
&& \qquad -
\frac{\omega}{2\pi} \, 2 \, f(t,\omega) \,  
\int_0^\infty  \omega' \, d\omega' \, 
 s^\smA(p; \omega',\omega) \,  f(t,\omega') \,.
\label{cccolll}
\end{eqnarray}

\section{Expansions and Angular Averages}
\label{outline}

It proves convenient to use the angles $\alpha$ , $\alpha'$ between 
${\bf p}$ and ${\bf k}$ , ${\bf k}'$, and the angle $\theta$ 
between ${\bf k}$ and ${\bf k}'$. We also use the velocity
\begin{equation}
v = \frac{|{\bf p}|}{E} = \frac{|{\bf p}|}{E(|{\bf p}|)} < 1 \,.
\end{equation}
The delta function in Eq.~(\ref{ssss}) now becomes
\begin{equation}
\delta \Big(  2 p \left( k - k' \right) - 2 k k' \Big) = 
\frac{1}{2 E} \ \delta \Big( 
\omega \left(1 - v \cos\alpha  \right) -
\omega' \left(1 - v \cos\alpha'  \right) 
- (\omega \omega'/ E)  \left( 1 - \cos\theta \right) \Big) \,,
\label{delta}
\end{equation}
and the squared scattering amplitude (\ref{amp2'}) 
now appears as
\begin{eqnarray}
    |T(p',k'; p,k)|^2 &=&   6\pi \, m_e^2 \, \sigma_\smT
   \, \Bigg\{  
  2  + \frac{1 - \cos\theta}{( 1 - v \cos\alpha)(1- v \cos\alpha')} 
     \left[\frac{\omega\omega'}{E^2} \, (1 - \cos\theta)
         - 2 \left( 1 - v^2\right) \right] 
\nonumber\\
&& \qquad\qquad
+ \left( 1 - v^2 \right)^2 \, 
\frac{(1 - \cos\theta)^2}{( 1 - v \cos\alpha)^2(1- v\cos\alpha')^2}  \Bigg\} \,.
\label{amp2aa}
\end{eqnarray}

As we shall see, the Kompaneets equation results from expanding 
Eqs.~(\ref{delta}) and (\ref{amp2aa}) in powers of $v$, with this
equation resulting from the order $v^2$ terms.  The leading relativistic
corrections that concern us require that the expansion be carried out
to order $v^4$. We note that since in the applications that we envisage,
$\omega ,\, \omega' \sim T \sim p^2 / m_e \sim v^2 \, E$,
 $\omega / E$ or $\omega' / E$
should be counted as being of order $v^2$. 

\newpage

The needed expansion of the delta function (\ref{delta}) in powers of 
$v$ reads
\begin{eqnarray}
2E\, \delta \Big(  2 p \left( k - k' \right) - 2 k k' \Big) &=& 
 \delta \Big( \omega - \omega' \Big) 
\nonumber\\
&-&
\Bigg[v \left(\omega \cos\alpha - \omega' \cos\alpha' \right)
\delta' \Big( \omega - \omega' \Big) \Bigg]
\nonumber\\
&+&
\Bigg[ \frac{1}{2} \, v^2\left(\omega \cos\alpha - \omega' \cos\alpha' \right)^2
\delta'' \Big( \omega - \omega' \Big) 
\nonumber\\
&& \qquad\qquad\qquad\qquad\qquad
- \frac{\omega \omega'}{E} \left( 1 - \cos\theta \right) 
   \delta'\Big( \omega - \omega' \Big) \Bigg]  
\nonumber\\
&-&
\Bigg[ \frac{1}{3!} \, v^3  \left(\omega \cos\alpha - \omega'\cos\alpha' \right)^3
\delta''' \Big( \omega - \omega' \Big) 
\nonumber\\
&&
- \, v \left(\omega \cos\alpha - \omega' \cos\alpha' \right)
\frac{\omega \omega'}{E} \left( 1 - \cos\theta \right) 
   \delta''\Big( \omega - \omega' \Big) \Bigg]  
\nonumber\\
&+& 
\Bigg[ \frac{1}{4!} \, v^4 \left(\omega \cos\alpha - \omega' \cos\alpha' \right)^4
\delta'''' \Big( \omega - \omega' \Big) 
\nonumber\\
&&
- \frac{1}{2}\, v^2 \left(\omega \cos\alpha - \omega' \cos\alpha' \right)^2
\frac{\omega \omega'}{E} \left( 1 - \cos\theta \right) 
   \delta'''\Big( \omega - \omega' \Big)
\nonumber\\
&& \qquad 
+ \frac{1}{2} \, \left(\frac{\omega \omega'}{E}\right)^2 \left( 1 - \cos\theta \right)^2 
   \delta''\Big( \omega - \omega' \Big) \Bigg] 
+ \, {\cal O}(v^5) \,.
\label{deltaa}
\end{eqnarray}
The term of order $v^0$, $\delta(\omega - \omega')$, makes
no contribution, since for it the two parts of the collision integral,
the `scattering in to' and the `scattering out of', cancel. Hence we 
need expand the squared amplitude (\ref{amp2aa}) only to order $v^3$ 
to obtain results good to order $v^4$:
\begin{eqnarray}
    |T(p',k'; p,k)|^2 &\simeq&   6\pi \, m_e^2 \, \sigma_\smT
   \Bigg\{ v^0 \, \Bigg[ \left(1 + \cos^2\theta \right) \Bigg]
\nonumber\\
&-& 2 \, v \, \left( 1 - \cos\theta \right) \, 
\Bigg[\left( \cos\alpha + \cos\alpha' \right) \, \cos\theta \Bigg]
\nonumber\\
&+& v^2 \, \left( 1 - \cos\theta \right) \,
\Bigg[\left( \cos^2\alpha + \cos^2\alpha' + 2 \cos\alpha \cos\alpha' \right)
\nonumber\\
&& \qquad
+ \left( 2 -3\cos^2\alpha -3\cos^2\alpha'
 -4 \cos\alpha \cos\alpha' \right) \cos\theta \Bigg]
\nonumber\\
&+& v^3 \, \left( 1 - \cos\theta \right) \,
 \Bigg[ \Big( - 2 \cos\alpha - 2 \cos\alpha' + 2 \cos^3\alpha + 2 \cos^3\alpha' 
\nonumber\\
&& \qquad\qquad\qquad   
 + 4\cos^2\alpha\cos\alpha' + 4\cos\alpha\cos^2\alpha' \Big)
\nonumber\\
&& \qquad
-  \Big(- 4 \cos\alpha - 4 \cos\alpha' + 4 \cos^3\alpha + 4 \cos^3\alpha' 
\nonumber\\
&& \qquad\qquad\qquad
 + 6 \cos^2\alpha\cos\alpha' + 6 \cos\alpha\cos^2\alpha' \Big) \cos\theta \Bigg]
  \Bigg\} \,.
\label{amp22a}
\end{eqnarray}

We now multiply the expressions (\ref{deltaa}) and (\ref{amp22a}) together
and retain the resulting terms up to those of order $v^4$. We then use  
the result of angular averaging over the directions of ${\bf p}$ detailed 
in Appendix \ref{angave} and subsequently average over the direction between 
${\bf k}$ and ${\bf k}'$, the average over $\cos\theta$. To present the
results in a compact form, we separately record the order $v^2$ result,
the one that gives the Kompaneets equation,
\begin{eqnarray}
s_2(p;\omega',\omega) &=&
 \langle |T(p',k'; p,k)|^2 
 \delta \Big(2p\left(k - k'\right)-2kk'\Big) \rangle\Big|_2 
\nonumber\\
&=& 
     4 \, \pi \,  \frac{m_e^2}{E} \, \sigma_\smT  \, \Bigg\{
 v^2 \, \frac{\omega\omega'}{3} \, \delta'' \Big( \omega - \omega' \Big)
 - \frac{\omega \, \omega'}{E} \,  
   \delta'\Big( \omega - \omega' \Big)  \Bigg\} \,,
\label{Tfor2}
\end{eqnarray}
and the order $v^4$ result which gives the leading relativistic correction
to the Kompaneets equation
\begin{eqnarray}
s_4(p;\omega',\omega) &=& \langle|T(p',k'; p,k)|^2 \,
      \delta \Big(2p\left(k - k'\right)-2kk'\Big) \rangle \Big|_4
\nonumber\\
&=&  
 \frac{\pi}{15} \, \frac{m_e^2}{E}  \, \sigma_\smT \,  
 \Bigg\{ v^4 \, \bigg[ 2 \left( \omega - \omega' \right)^2 \omega \, \omega'
+ \frac{14}{5} \, \omega^2 \, \omega^{\prime \, 2} \Bigg] \, 
\delta'''' \Big( \omega - \omega' \Big) 
\nonumber\\
&&
   -  
 10 \, v^2 \, \Bigg[ \frac{2 }{25} \, v^2 
    \,   \left( \omega - \omega' \right) \omega \, \omega'
       +  \left(  \left(\omega  - \omega' \right)^2 
         + \frac{14}{5} \omega \, \omega' \right) \, 
          \frac {\omega \, \omega'}{E} \Bigg] \,
   \delta'''\Big( \omega - \omega' \Big)
\nonumber\\
&& 
+ 
\Bigg[ -\frac{16}{5} \, v^4 \, \omega \, \omega' +
 4 \, v^2 \,  \left(\omega - \omega' \right) 
          \frac {\omega \, \omega'}{E} 
    + 42  \left(\frac {\omega \, \omega'}{E} \right)^2  \Bigg] \,  
\delta'' \Big( \omega - \omega' \Big) 
\nonumber\\
&& +
 \Bigg[ 12 \, v^2 \, \frac{\omega \, \omega'}{E}
            \Bigg] \,  \delta'\Big( \omega - \omega' \Big) \Bigg\} \,.
\label{Tfor4}
\end{eqnarray}
There are no terms of odd order in $v$ in the somewhat lengthly algebra
required to obtain these formulae. They simplify with the aid of the 
delta function identities presented in Appendix \ref{dids}:
\begin{eqnarray}
\frac{\omega \, \omega'}{2\pi \, m_e^2} \, s^\smS_2(p;\omega',\omega) &=&
 2 \, \sigma_\smT \, v^2 \, \frac{ (\omega \, \omega')^2}{3 \, E} \,
 \delta'' \Big( \omega - \omega' \Big) \,,
\label{ssfor2}
\end{eqnarray}
\begin{eqnarray}
\frac{\omega \, \omega'}{2\pi \, m_e^2} \, s^\smA_2(p;\omega',\omega) &=&
   - 2 \, \sigma_\smT \, \frac{(\omega \, \omega')^2}{E^2} \,
   \delta'\Big( \omega - \omega' \Big)  \,,
\label{safor2}
\end{eqnarray}
and
\begin{eqnarray}
\frac{\omega \, \omega'}{2 \pi \, m_e^2} \,s^\smS_4(p;\omega',\omega) 
&=& 
\sigma_\smT \, \frac{(\omega \, \omega')^2}{15 \, E} \Bigg\{
  \frac{7}{5} \,  v^4 \, \omega \, \omega' \,
\delta''''\left( \omega - \omega' \right)
\nonumber\\
&& \qquad\qquad\qquad\quad + \Bigg[ \frac{58}{5} \, v^4 + 
21 \, \frac{\omega \, \omega'}{E^2} \Bigg]
\delta''\left( \omega - \omega' \right) \Bigg\} \,,
\label{ssfor4}
\end{eqnarray}
\begin{eqnarray}
\frac{\omega \, \omega'}{2 \pi\, m_e^2} \,s^\smA_4(p;\omega',\omega) 
&=&  
   - \sigma_\smT \, v^2 \,
    \frac {(\omega \, \omega')^2}{15 \, E^2} 
        \Bigg\{ 14 \, \omega \, \omega' \, \delta'''\Big( \omega - \omega' \Big) 
+ 28 \, \delta'\left(\omega  - \omega' \right) \Bigg\} \,.
\label{safor4}
\end{eqnarray}

\section{Results}
\label{results}

It is now a straightforward although tedious matter to insert the
forms above into Eq.~(\ref{cccolll}), perform the $\omega'$ integrals,
and place the results into Eq.~(\ref{colll}) to secure the Kompaneets
equation and its leading relativistic corrections.

\subsection{General Result}
\label{A}

\begin{eqnarray}
\omega^2 \, \frac{\partial}{\partial t} \, f(t,\omega)\
&=&
  \sigma_\smT \, \frac{n_e}{ m_e \, c}  \, 
 \frac{d}{d\omega} \, \omega^4 \,\Bigg\{
    T_{\rm eff \, 1}  \, \frac{d \, f(t,\omega)}{d\omega} \,
 + \hbar \,\left[ 1 + f(t,\omega) \right] \, f(t,\omega) \ 
\nonumber\\
&+&         
 \frac{1}{m_e \, c^2} \left( \frac{5}{2} \, T^2_{\rm eff \, 2} 
+ 21 \,T^2_{\rm eff \, 1} \right)  \,  \frac{d \, f(t,\omega)}{d\omega}
+ \frac{47}{2 m_e \, c^2 } \, T_{\rm eff \, 1} \, 
\hbar \, \left[ 1 + f(t,\omega) \right] \, f(t,\omega) 
\nonumber\\
&-&  \frac{7 \, \hbar \omega^2}{10 m_e \, c^2}  \, \Bigg[
 6 \, T_{\rm eff \, 1} \, \left( \frac{d \, f(t,\omega) }{d\omega}  \right)^2  
     - \hbar \frac{d \, f(t,\omega)}{d\omega} +
 T^2_{\rm eff \, 1} \frac{1}{\hbar} \, \frac{d^3 f(t,\omega)}{d\omega^3} \Bigg] \Bigg\}
\nonumber\\
&-&  \frac{21 \,n_e \, \sigma_\smT }{5 \, m^2_e \, c^3}   
 \frac{d^2}{d\omega^2} \, \omega^5  \Bigg\{
\Big( T^2_{\rm eff \, 2} + 2 \,T^2_{\rm eff \, 1} \Big) 
\, \frac{d \, f(t,\omega)}{d\omega}
+ 3  T_{\rm eff \, 1} \, \hbar  \left[ 1 + f(t,\omega) \right]  f(t,\omega) \bigg\}
\nonumber\\
&+& \frac{7 \, n_e \, \sigma_\smT}{10 m^2_e \, c^3}  
 \frac{d^3}{d\omega^3} \, \omega^6 \,\Bigg\{
 \Big( T^2_{\rm eff \, 2} + T^2_{\rm eff \, 1} \Big) 
 \frac{d \, f(t,\omega)}{d\omega}
+2  T_{\rm eff \, 1} \, \hbar  \left[ 1 + f(t,\omega) \right]  f(t,\omega) \Bigg\} \,.
\nonumber\\
&&
\label{whee!}
\end{eqnarray}
Here we have reverted to conventional units and used the effective temperature 
definitions (\ref{Tone}) and (\ref{Ttwo}) in Appendix \ref{effect} which, for 
convenience, we repeat here:
\begin{equation}
 T_{\rm eff \, 1} = \frac{1}{ n_e}  \, 
\int \frac{(d^3{\bf p})}{(2\pi)^3} \, \frac{p^2 \, c^2}{3 E(p)} \, 2 \, g(t,E) \,,
\label{T1}
\end{equation}
and
\begin{equation}
 T^2_{\rm eff \, 2} = \frac{4}{15 \, n_e} \,
\int \frac{(d^3{\bf p})}{(2\pi)^3} \, \left(\frac{p^2}{2m_e} \right)^2 \, 2 \, 
    g(t,E) \,.  
\label{T2} 
\end{equation} 

We have chosen to order the terms in our result (\ref{whee!}) so as to
have successive parts involve overall higher derivatives. Since each
part starts out with at least one overall derivative, the result
conserves photon number as it must. Only the part with the single
overall derivative $d/d\omega$ contributes to the rate of energy
exchange between the photons and the electrons. Similarly, the rate at
which the second moment $(\hbar\omega)^2$ changes with time is
affected only by the parts involving $d/d\omega$ and $d^2/d\omega^2$
while all the parts contribute to the time rate of change of the
$(\hbar\omega)^3$ moment. We have kept some photon frequency
derivatives within the sequence of increasingly higher overall
derivative so that the sum of the terms in each of these groups vanishes
in thermal equilibrium. Each of these groupings in the result
(\ref{whee!})  vanishes, in fact, in the more general situation in
which the photon distribution is of the Planck form (\ref{Planck}) but with
the electron $g(E)$ constrained only to have 
$T_{\rm eff \, 1} =T_{\rm eff \, 2} = T_\gamma$. 
With the electrons in thermal equilibrium, $ T_{\rm eff \, 1} $ 
reduces without approximation to the electron temperature $T_e$, and 
$T_{\rm eff \, 2} \to T_e$ in first approximation, an approximation 
that suffices since $ T_{\rm eff \, 2}$ only occurs in the relativistic 
correction terms. In this case, each of these groupings of terms vanishes 
for a Planck distribution when  $T_\gamma = T_e$.  Our expression (\ref{whee!})
is in complete agreement with the work of Challinor and Lasenby \cite{C&L} 
in  the limit in which the electrons are in thermal equilibrium at temperature 
$T_e$. However, as mentioned previously, the structure of 
our expression (\ref{whee!}) for electrons in thermal equilibrium --- which 
is equivalent to that of C\&L --- differs completely from the previous 
(incorrect) results of Cooper \cite{coop} and of Prasad, Shestakov, 
Kershaw, and Zimmerman \cite{prasad}.

\subsection{Photons In Thermal Equilibrium}
\label{B}

For photons in thermal equilibrium at a temperature $T_\gamma$ [the Planck 
distribution (\ref{pno}) or, more generally, a Bose-Einstein distribution of photons]
the result reduces to
\begin{eqnarray}
\omega^2 \, \frac{\partial}{\partial t} \, f(t,\omega)\
&=& 
  - 4 \, \frac{d}{d\omega} \, \omega^3 \, \sigma_\smT \, \frac{n_e}{ m_e \, c}  \, 
\Bigg\{ \left(1 - \frac{21}{20} \, \frac{\hbar^2\, \omega^2}{m_e c^2 \,  T_\gamma} \right) \,
 \Big( T_{\rm eff \, 1} - T_\gamma \Big)
\nonumber\\
&& \qquad\qquad\qquad\qquad\qquad       
+ \frac{5}{2\, m_e \, c^2}  \, \Big( T^2_{\rm eff \, 2} - T_{\rm eff \, 1} \, T_\gamma \Big)
 \,  \Bigg\} \, f_0(\omega)
\nonumber\\
&&
+  \,  \frac{d^2}{d\omega^2} \, \omega^4  \, \sigma_\smT \, \frac{ n_e}{ m_e \, c}  \, 
\Bigg\{  \left(1 - \frac{7}{10} \, \frac{\hbar^2 \, \omega^2}{m_e c^2\, T_\gamma} \right) \,
 \Big( T_{\rm eff \, 1} - T_\gamma \Big)
\nonumber\\
&& \qquad\qquad\qquad\qquad\qquad\,\,
+ \frac{47}{2 \, m_e \, c^2} \Big(T^2_{\rm eff \, 2} - T_{\rm eff \, 1} \, T_\gamma \Big) 
          \Bigg\} \, f_0(\omega)
\nonumber\\
&& - \,
 \frac{d^3}{d\omega^3} \, \omega^5  \,  \sigma_\smT \, \frac{ 42 \,n_e}{5 \, m^2_e \, c^3} 
\,  \Big( T^2_{\rm eff \, 2}  -  \, T_{\rm eff \, 1} \, T_\gamma \Big) \, f_0(\omega)
 \nonumber\\
&&+ \,
 \frac{d^4}{d\omega^4} \, \omega^6  \,
  \sigma_\smT \, \frac{7 \, n_e}{10 \, m^2_e \, c^3}  \, 
 \Big( T^2_{\rm eff \, 2} - T_{\rm eff \, 1} \, T_\gamma \Big) \, 
  f_0(\omega) \,.
\label{whiz}
\end{eqnarray}

\subsection{Energy Transport}
\label{C}

The energy transfer per unit volume to the photons is given by
\begin{equation}
\dot u_\gamma = 2 \int \frac{(d^3{\bf k})}{(2\pi)^3} \, \omega \, 
\frac{\partial}{\partial t} \, f(t,\omega) \,.
\end{equation}
From Eq.~(\ref{whiz}), we see that, including the first relativistic
corrections, this energy transfer between a photon distribution in
equilibrium at temperature $T_\gamma$ and an arbitrary isotropic
distribution of election energies involves
\begin{eqnarray}
2 \int \frac{(d^3{\bf k})}{(2\pi)^3} \, \omega \, f_0(\omega)
&=& \sum_{n=1}^{\infty} \, \frac{1}{\pi^2} \, \int_0^\infty d\omega 
\, \omega^3 \, \exp\left\{ - n \, \frac{\omega}{T_\gamma} \right\}
\nonumber\\
&=& T^4_\gamma \, \frac{3!}{\pi^2} \sum_{n=1}^{\infty} \, \frac{1}{n^4} = 
 T^4_\gamma \, \frac{3!}{\pi^2} \, \zeta(4) = 
   \frac{ \pi^2}{15} \, T^4_\gamma = u_\gamma \,,
\end{eqnarray}
and
\begin{eqnarray}
2 \int \frac{(d^3{\bf k})}{(2\pi)^3} \, \omega^3 \, f_0(\omega)
&=& \sum_{n=1}^{\infty} \, \frac{1}{\pi^2} \, \int_0^\infty d\omega 
\, \omega^5 \, \exp\left\{ - n \, \frac{\omega}{T_\gamma} \right\}
\nonumber\\
&=& T^6_\gamma \, \frac{5!}{\pi^2} \sum_{n=1}^{\infty} \, \frac{1}{n^6} = 
 T^6_\gamma \, \frac{5!}{\pi^2} \,\zeta(6) = 
   \frac{8 \, \pi^4}{63} \, T^6_\gamma  
= \frac{120 \, \pi^2}{63} \, T^2_\gamma \, u_\gamma\,.
\end{eqnarray}
Hence, again reverting to conventional units, 
\begin{eqnarray}
\dot u_\gamma =  4 \, \sigma_\smT \, \frac{n_e}{ m_e \, c}  \, 
\Bigg\{ \left(1 - 2 \, \pi^2 \, \frac{T_\gamma}{m_e \, c^2} \right) \,
 \Big( T_{\rm eff \, 1} - T_\gamma \Big) 
+ \frac{5}{2\, m_e \, c^2}  \,  
\Big( T^2_{\rm eff \, 2} - T_{\rm eff \, 1} \, T_\gamma \Big) \,  \Bigg\} \, 
 u_\gamma \,.
\label{Erate}
\end{eqnarray}

In the electron thermal equilibrium limit in which 
$T_{\rm eff \, 1} =T_{\rm eff \, 2} = T_e$, 
the energy transfer rate (\ref{Erate}) agrees
with the rate given by Woodward \cite{wood} which was later confirmed
by Challinor and Lasenby \cite{C&L}. The result (\ref{Erate}) holds,
of course, for a general isotropic distribution of electrons with the
two effective temperatures $T_{\rm eff \, 1}$ and $T_{\rm eff \, 2}$
defined by the integrals (\ref{T1}) and (\ref{T2}).

\subsection{Sunyaev-Zel'dovich Effect For Non-thermal Electrons}
\label{D} 

Equation (\ref{whiz}) can be used to generalize published results on
the Sunyaev-Zel'dovich effect, the distortion of the cosmic microwave
background, a Planck distribution at a very low temperature, by high
energy electrons in hot plasmas in galactic clusters.  The distortion
involves changing the time derivative in the Boltzmann equation to the
proper coordinate distance $\ell$ along the line of sight through the
plasma cloud, $t \to \ell / c$.  In doing so in the relativistic
corrected Kompaneets formula (\ref{whiz}), we encounter two
dimensionless variables
\begin{equation}
y_1 = \sigma_\smT \, \int d\ell \, n_e(\ell) \, 
      \frac{T_{\rm eff \, 1}(\ell)}{m_e \, c^2 } \,,
\end{equation}
and
\begin{equation}
y_2 = \sigma_\smT \, \int d\ell \, n_e(\ell) \, 
       \frac{T^2_{\rm eff \, 2}(\ell)}{m^2_e \, c^4 } \,,
\end{equation}
that replace the conventional $y$ parameter\footnote{Note that even 
in the case of a plasma in local thermodynamic equilibrium with an 
electron temperature $T_e(\ell)$, the relativistic corrections involve 
a different $y$ parameter ($y = y_2$) defined by the electron number 
weighted average of $T^2_e(\ell)$ rather than the first power 
$T_e(\ell)$ that appears in $y = y_1$.}. It is now convenient to define
\begin{equation} 
x = \frac{\hbar \omega}{T_\gamma} \,,
\end{equation}
and write the Planck distribution as 
$f_0(x) = [\exp\{x\} - 1 ]^{-1}$.  Since the 
microwave background temperate $T_\gamma$ is so low, while the electrons have 
average energies that are several keV, we may neglect the very small ratios 
$T_\gamma / T_{\rm{eff} \, 1} $ and $T_\gamma / T_{\rm{eff} \, 2}$ as well as 
$T_\gamma / m_e c^2$. With the omission of these  terms, carrying out the 
derivatives in Eq.~(\ref{whiz}) yields the spectral distortion
\begin{eqnarray}
\frac{\Delta f(x)}{f_0(x)} &=& y_1 \, x \, \left[ 1 + f_0(x) \right] \,
\Big\{ x - 4 + 2 \, x f_0(x) \Big\} 
\nonumber\\
&& +
 y_2 \, x \, \left[ 1 + f_0(x) \right] \, \Big\{
  \frac{x}{10}  \left( 235 - 84\,x+
7\,x^2 \right) - 10 
\nonumber\\
&& \qquad\quad +
 \frac{1}{5} \, \left(235 -252\,x + 49\,x^2 \right)  x \, f_0 (x) 
\nonumber\\
&& \qquad\quad +
 \frac{126}{5} \, (x-2) \, x^2 f_0^2(x) + \frac{84}{5} \, x^3 f_0^3 (x) \Big\} \,.
\label{SZ}
\end{eqnarray}
In the limit of small $x$, the Rayleigh-Jeans region, Eq.~(\ref{SZ}) simplifies to
\begin{equation}
\frac{\Delta f(x)}{f_0(x)} = - 2 \, y_1 + \frac{17}{5} \, y_2 \,.
\label{RJ}
\end{equation}
In contrast to previous papers on the Sunyaev-Zel'dovich effect, Eqs.~(\ref{SZ}) 
and (\ref{RJ}) hold when the electrons are not in thermal equilibrium; when they are in 
equilibrium Eqs.~(\ref{SZ}) and (\ref{RJ}) agree with the expressions obtained by Challinor 
and Lasenby\footnote{These authors, however, do not define the proper parameters
$y_1$ and $y_2$ that, as noted in the previous footnote, are needed for the
relativistic treatment, but rather use a $y=y_1$ and then multiply this by an undefined
electron temperature to obtain a $y_2$ parameter for the relativistic corrections.} \cite{C&L}.

\newpage

\appendix

\section{Squared Amplitude Details}
\label{details}

We first express the fully relativistic squared amplitude \cite{text} as
\begin{eqnarray}
    |T(p',k'; p,k)|^2 =  6  \pi \, m_e^2 \, \sigma_\smT \Bigg\{  
   \left( \frac{\kappa'}{\kappa} + \frac{\kappa}{\kappa'} \right)
+ 2 \left( \frac{m^2_e}{\kappa}  - \frac{m^2_e}{\kappa'} \right)
    + \left( \frac{m^2_e}{\kappa}  - \frac{m^2_e}{\kappa'} \right)^2
      \Bigg\} \,,
\label{amp2}
\end{eqnarray}
in which $ \sigma_\smT = 8\pi \, r_0^2 / 3$ is the Thomson cross section with 
$r_0$ the classical electron radius and, with our space-like metric 
\begin{equation}
\kappa = - pk = p^0 k^0 - {\bf p} \cdot {\bf k}
 \ \,, \qquad\qquad \kappa' = - pk' = p^0 {k'}^0 - {\bf p} \cdot {\bf k}' \,. 
\end{equation}
Note that in terms of these variables the delta function constraint
(\ref{delta}) reads
\begin{equation}
\kappa' - \kappa = k k' \,.
\label{relate}
\end{equation}
It is convenient to use the variable 
\begin{equation}
\bar\kappa = \sqrt{\kappa\kappa'} \,,
\end{equation}
so that the relation (\ref{relate}) may be written as
\begin{equation}
\frac{\kappa'}{\kappa} = 1 + \frac{k k'}{\kappa} 
      = 1 + \sqrt{\frac{\kappa'}{\kappa}} \, \frac{k k'}{\bar\kappa} \,.
\end{equation}
The proper solution of this quadratic equation, written in terms of 
$\kappa'/\kappa$, is
\begin{equation}
\frac{\kappa'}{\kappa} = 
    \frac{1}{2} \left[ \left(\frac{k k'}{\bar\kappa}\right)^2 + 2 
  +\frac{k k'}{\bar\kappa} \sqrt{ \left(\frac{k k'}{\bar\kappa}\right)^2 + 4} 
\,\, \right] \,.
\end{equation}
Similarly,
\begin{equation}
\frac{\kappa}{\kappa'} = 
    \frac{1}{2} \left[ \left(\frac{k k'}{\bar\kappa}\right)^2 + 2 
  - \frac{k k'}{\bar\kappa} \sqrt{ \left(\frac{k k'}{\bar\kappa}\right)^2 + 4} 
\,\, \right] \,.
\end{equation}
Making use of  Eq.~(\ref{relate}) and some algebra now presents
\begin{eqnarray}
    |T(p',k'; p,k)|^2 &=&  6\pi \, m_e^2 \, \sigma_\smT \,  \left\{  
 \left(\frac{k k'}{\bar\kappa}\right)^2 + 2 
 + \,  2 \, m^2_e \, \frac{k k'}{\bar\kappa^2} 
  \, + \, m^4_e \, \left( \frac{k k'}{\bar\kappa^2} \right)^2  \right\} \,.
\label{amp2'}
\end{eqnarray}

\section{Angular Averages}
\label{angave}

The calculation outlined in the text involves the integration of
\begin{equation} 
\cos\alpha = \hat{\bf p} \cdot \hat{\bf k} = \hat
p^{\,l} \, \hat k^{\,l} \,, \qquad\qquad \cos\alpha' = \hat{\bf p}
\cdot \hat{\bf k}' = \hat p^{\,l} \, \hat k^{\prime \, l} \,,
\end{equation} 
and of the products of the powers $\cos^m\alpha
\, \cos^n\alpha'$, over the solid angle of the momentum ${\bf p}$
associated with the electron distribution $g(t,E)$. This is equivalent
to averaging over all orientations of the unit vector $\hat{\bf p}$,
and this averaging can be performed at any stage of the computation
--- it may be performed before the actual integral over ${\bf p}$ is
carried out. The averages may be expressed as contractions of outer
products $\hat k^l \, \cdots$ and $\hat k^{\prime \, m} \cdots$ with the
rotationally invariant tensors that result from the angular averages 
$\langle \hat p^k \, \hat p^l \, \cdots \rangle$.  For example,
\begin{equation}
\langle \cos\alpha \, \cos\alpha' \rangle = 
\hat k^l \, \hat k^{\prime \, m} \, \langle \hat p^k \, \hat p^m \rangle \,.
\end{equation}
This is the method that we shall employ.

Under the  angular average
\begin{equation}
\langle \hat p^{\, l} \rangle = 0 \,,
\end{equation}
and since ${\bf p}$ is a vector, not a pseudo-vector,
\begin{equation}
\langle \hat p^{\,l} \, \hat p^{\,m} \, \hat p^{\,n} \rangle = 0 \,,
\end{equation}
because the only rotationally invariant, third rank tensor, is the 
pseudo-tensor $\epsilon^{lmn}$. The lowest-order correlation is
\begin{equation}
\langle \hat p^{\,l} \hat p^{\,m} \rangle = \frac{1}{3} \, \delta^{lm} \,,
\label{two}
\end{equation}
where $\delta^{kl}$ is the matrix element of the invariant unit matrix. The 
overall coefficient is determined by the trace
$
 \langle \hat p^{\,l} \hat p^{\,l} \rangle = \langle 1 \rangle = 1 \,.
$
The final average that we shall need is
\begin{equation}
\langle \hat p^{\,k} \hat p^{\,l} \hat p^{\,m} \hat p^{\,n} \rangle 
= \frac{1}{15} \, \left[ \delta^{kl} \, \delta^{mn} + \delta^{km} \, \delta^{ln} 
     + \delta^{kn} \, \delta^{lm} \right] \,.
\label{four}
\end{equation}
Here the particular combination of the delta symbols on the right-hand side 
is required to reproduce the complete symmetry of the left-hand side 
under any permutation of the indices $k\,,\,l\,,\,m\,,\,n$. The overall 
coefficient again follows from taking the trace over any index pair and 
comparing the result with Eq.~(\ref{two}).

Therefore,
\begin{equation}
\langle \cos\alpha \rangle = 0 = \langle \cos\alpha' \rangle \,,
\end{equation}
and
\begin{equation}
\langle \cos\alpha \cos\alpha' \rangle = \frac{1}{3} \, \hat{\bf k} \cdot \hat{\bf k}' 
                                       = \frac{1}{3} \, \cos\theta \,,
\end{equation}
from which follow
\begin{equation}
\langle \cos^2\alpha \rangle = \frac{1}{3} = \langle \cos^2\alpha' \rangle \,.
\end{equation}
Since the angular average of three momentum vectors vanishes,
\begin{eqnarray}
0 &=& \langle \cos^3\alpha \rangle = \langle \cos^2\alpha\cos\alpha' \rangle 
  = \langle \cos\alpha\cos^2\alpha'\rangle = \langle \cos^3\alpha \rangle \,. 
\end{eqnarray}
Next,
\begin{equation}
\langle \cos^3\alpha \cos\alpha' \rangle = \frac{1}{15} \left[ 
  \hat{\bf k}^2 \, \hat{\bf k} \cdot \hat{\bf k}' 
  + \hat{\bf k}^2 \, \hat{\bf k} \cdot \hat{\bf k}' 
  + \hat{\bf k} \cdot \hat{\bf k}' \, \hat{\bf k}^2 \right] 
   = \frac{1}{5} \, \cos\theta \,,
\end{equation}
and 
\begin{equation}
\langle \cos\alpha \cos^3\alpha' \rangle  = \frac{1}{5} \, \cos\theta \,.
\end{equation}
Finally
\begin{equation}
\langle \cos^2\alpha \cos^2\alpha' \rangle = \frac{1}{15} \left[ 
  \hat{\bf k}^2 \, \hat{\bf k}^{\prime \, 2} + (\hat{\bf k} \cdot \hat{\bf k}' )^2 
  + (\hat{\bf k} \cdot \hat{\bf k}' )^2 \right] 
   = \frac{1}{15} \left[ 1 + 2 \cos^2\theta \right] \,,
\end{equation}
from which follow
\begin{equation}
\langle \cos^4\alpha \rangle = \frac{1}{5}  = \langle \cos^4\alpha' \rangle \,.
\end{equation}

\section{Delta Function Identities}
\label{dids}

The work in the text involves various derivatives of 
$\delta(\omega -\omega') = \delta(x)$ multiplied by various 
powers of $\omega - \omega' = x$. Simple manipulations 
can be performed to place a derivative
$d^n / dx^n$ to the left of a power $x^m$, leaving lower derivatives
and lower powers of $x$.  Again, the lower derivatives can be ordered
to the left so that all the derivatives appear as total derivatives.
According to the rules of generalized functions, any resulting term of
the form $x^l \, \delta(x)$ gives a vanishing contribution.  The
simplest example of this procedure is
\begin{eqnarray}
x \, \delta'( x )
&=& \frac{d}{dx} \, \left[x \ \delta( x ) \right]
                    - \delta\left( x \right) \,.
\label{example}
\end{eqnarray}
As we have just noted, the first term may be discarded.  Moreover, the
delta function $\delta(x) = \delta(\omega - \omega')$ with no
derivative can also be omitted because it gives rise to equal contributions
from the `scattering in to' and `scattering out of' terms in the 
Boltzmann equation which cancel. We use the symbol $ \,\,`\!\!='\,$ to
denote the only effective parts that remain after the manipulations described
above have been made. Thus, we write Eq.~(\ref{example}) as
\begin{eqnarray}
x \, \delta'( x ) \,\,`\!\!='\,   0 \,. \null\qquad
\label{1}
\end{eqnarray}
More involved computations lead to the effective results
\begin{eqnarray}
x \, \delta''( x )  \,\,`\!\!='\,  - 2 \, \delta'( x ) \,,
\label{2}
\end{eqnarray}
\begin{eqnarray}
   x^2 \, \delta''( x )  \,\,`\!\!='\,  0 \,, \null\qquad
\label{3}
\end{eqnarray}
\begin{eqnarray}
   x \, \delta'''( x )  \,\,`\!\!='\,  - 3 \, \delta''( x ) \,,
\label{4}
\end{eqnarray}
\begin{eqnarray}
    x^2 \, \delta'''( x )  \,\,`\!\!='\,  6 \, \delta'( x ) \,,
\label{5}
\end{eqnarray}
\begin{eqnarray}
    x^2 \, \delta''''( x ) \,\,`\!\!='\,    + 12 \, \delta''( x ) \,,
\label{6}
\end{eqnarray}
and
\begin{eqnarray}
   x^4  \delta''''( x )  \,\,`\!\!='\,   0 \,. \null\qquad
\label{7}
\end{eqnarray}

\section{Effective Temperatures Defined By Electron Distribution Integrals}
\label{effect}

Here we shall explain the definitions of the two effective temperatures 
$T_{\rm eff \, 1}$ and $T_{\rm eff \, 2}$ that reduce to the electron 
temperature $T_e$ when the electron relativistic phase-space distribution 
$g(t,E)$ is restricted to be a Maxwell-Boltzmann distribution.

For our system of
free, relativistic electrons, the number density is given for an arbitrary
phase space distribution $g(t,E)$ by
\begin{equation}
n_e  =  \int \frac{(d^3{\bf p})}{(2\pi)^3} \, 2 g(t,E) \,.
\label{ne}
\end{equation}
For the case of thermal equilibrium,
\begin{equation}
g(p) = {\rm const.} \, \exp\left\{ - \frac{E(p)}{T} \right\} \,,
\end{equation}
in which 
\begin{equation}
E(p) = \left[ p^2 + m_e^2 \right]^{1/2}
\end{equation}
is the total relativistic energy of an electron with momentum $p$.

To obtain the definition of $T_{\rm eff \, 1}$, we note that
partial integration gives\footnote{What follows is a proof that, for a dilute, 
non-interacting gas, the familiar equation of state
$ p = n \, T$ holds even for relativistic particles.}
\begin{eqnarray}
n_e &=&  \int \frac{(d^3{\bf p})}{(2\pi)^3} \, 2 \, {\rm const.} \,
      \exp\left\{ - \frac{E(p)}{T} \right\} \, 
       \frac{1}{3} \, \frac{\partial}{\partial {\bf p}} \cdot {\bf p}
\nonumber\\
    &=& - \int \frac{(d^3{\bf p})}{(2\pi)^3} \, {\rm const.} \, 
\frac{1}{3} \, {\bf p} \cdot \frac{\partial}{\partial {\bf p}} \,
    2 \,  \exp\left\{- \frac{E(p)}{T} \right\} 
\nonumber\\
    &=& \frac{1}{T} \, \int \frac{(d^3{\bf p})}{(2\pi)^3} \, \frac{p^2}{3E(p)} \,
    2 \, g(t,E) \,.
\end{eqnarray}
Thus we define
\begin{equation}
 T_{\rm eff \, 1} = \frac{1}{ n_e}  \, 
\int \frac{(d^3{\bf p})}{(2\pi)^3} \, \frac{p^2}{3 E(p)} \, 2 \, g(t,E) \,.
\label{Tone}
\end{equation}
We have just shown that in thermal equilibrium, $T_{\rm eff \, 1} \to T$
is an exact, relativistic result.

To obtain the definition of $T_{\rm eff \, 2}$, we note
that we are computing only the first relativistic corrections to the Compton
Fokker-Planck equation. Hence, for an equilibrium distribution, 
we may approximate a relativistic correction integral by
\begin{eqnarray}
\int \frac{(d^3{\bf p})}{(2\pi)^3} \,  \left( \frac{p^2}{2m} \right)^2 
 \, 2 \, g(t,E)
&\simeq&
 2 \,  {\rm const.} \, 
\exp\left\{- m_e / T \right\}  \int \frac{(d^3{\bf p})}{(2\pi)^3} \, 
\left( \frac{p^2}{2m_e} \right)^2 \,\exp\left\{ - \frac{p^2}{2m_e T} \right\} 
\nonumber\\
 &\simeq&
 \frac{15}{4} \, T^2 \, n_e \,.
\end{eqnarray}
Therefore, for an arbitrary distribution, we shall define 
\begin{equation}
 T^2_{\rm eff \, 2} = \frac{4}{15 \, n_e} \,
\int \frac{(d^3{\bf p})}{(2\pi)^3} \, \left(\frac{p^2}{2m_e} \right)^2 \, 2 \, 
    g(t,E) \,.
\label{Ttwo}
\end{equation}
To leading order, $T_{\rm eff \, 2} \to T$ when $g(t,E)$ becomes a thermal 
distribution.

With these results in hand, we can now evaluate the the integrals needed in
the text.  The following two integrals appear with the lowest-order functions
$s_2^\smS$ and $s_2^\smA$, and thus they must be evaluated to both lowest and
first non-leading orders:
\begin{eqnarray}
\int \frac{(d^3{\bf p})}{(2\pi)^3} \, \frac{v^2}{E^2} \,  g(t,E) &=&
\int \frac{(d^3{\bf p})}{(2\pi)^3} \, \frac{p^2}{E^4} \,  g(t,E) 
\nonumber\\
&\simeq& \int \frac{(d^3{\bf p})}{(2\pi)^3} \, \frac{1}{m_e^3}
\left[ \frac{p^2}{E}  - \frac{6}{m_e} \, \left(\frac{p^2}{2m_e} \right)^2 \right] 
\,g(t,E)
\nonumber\\
&=& \frac{3}{2 } \, \frac{n_e}{m_e^3} \, 
 \,  \left[  \, T_{\rm eff \, 1} \, 
- \frac{15}{2} \, \frac{T^2_{\rm eff \,2}}{m_e} \right] \,,
\end{eqnarray}
\begin{eqnarray}
\int \frac{(d^3{\bf p})}{(2\pi)^3} \, \frac{1}{E^3} \, g(t,E)
 &\simeq&
\int \frac{(d^3{\bf p})}{(2\pi)^3} \, \frac{1}{m_e^3} 
  \left[ 1 - \frac{3}{2} \, \frac{p^2}{m_e^2} \right] \, g(t,E)
\nonumber\\
  &\simeq&
\frac{n_e}{2 m_e^3} 
\left[ 1 - \frac{9}{2} \, \frac{T_{\rm eff \, 1}}{m_e} \right] \,.
\end{eqnarray}
On the other hand, for the higher order $s_4^\smS$ and $s_4^\smA$ terms, we
need only leading evaluations:
\begin{equation}
\int \frac{(d^3{\bf p})}{(2\pi)^3} \, \frac{v^4}{E^2} \, g(t,E) \simeq
\frac{15}{2} \, \frac{n_e}{m_e^2} \, \frac{T^2_{\rm eff \, 2}}{m_e^2} \,,
\end{equation}
\begin{equation}
\int \frac{(d^3{\bf p})}{(2\pi)^3} \, \frac{1}{E^4} \,  g(t,E) \simeq
\frac{n_e}{2 m_e^4} \,,
\end{equation}
\begin{equation}
\int \frac{(d^3{\bf p})}{(2\pi)^3} \, \frac{v^2}{E^3} \,  g(t,E) \simeq
\frac{3}{2} \, \frac{n_e}{m_e^4} \, T_{\rm eff \, 1} \,.
\end{equation}

\end{document}